\documentclass[conference]{IEEEtran} 
\usepackage{amsmath}
\usepackage{amsmath,amssymb,amsfonts}
\usepackage{cite}
\usepackage{mathtools}
\usepackage[binary-units=true]{siunitx}
\DeclareSIUnit{\belmilliwatt}{Bm}
\usepackage{amsfonts}
\usepackage{amssymb}
\usepackage{graphicx}
\usepackage{dsfont}
\usepackage{subcaption}
\usepackage{comment}
\usepackage{algorithm}
\usepackage{algpseudocode}
\usepackage{mathtools}

\usepackage{textcomp}
\usepackage[export]{adjustbox}
\usepackage{float}
\usepackage{booktabs}
\usepackage{multicol}
\usepackage{xparse}
\usepackage[left=1.62cm,right=1.62cm,top=1.9cm]{geometry}
\usepackage{color,soul}
\usepackage[bottom]{footmisc}
\usepackage[binary-units=true]{siunitx}
\DeclareSIUnit{\belmilliwatt}{Bm}
\DeclareSIUnit{\dBm}{\deci\belmilliwatt}
\DeclareSIUnit[per-mode=symbol,per-symbol=p]{\Bps}{\byte\per\second}

\sethlcolor{yellow}
\usepackage{algpseudocode}
\DeclareMathOperator{\E}{\mathbb{E}}
\usepackage{hyperref}

\makeatletter
\def\BState{\State\hskip-\ALG@thistlm}
\makeatother

\newcommand{\CS}[1]{\textcolor{magenta}{#1}}

\IEEEoverridecommandlockouts
\begin{document}

    \title{Timely and Efficient Information Delivery in  Real-Time Industrial IoT Networks}

\author{
\IEEEauthorblockN{Hossam Farag\IEEEauthorrefmark{1}, Dejan Vukobratovi\' c\IEEEauthorrefmark{2}, Andrea Munari\IEEEauthorrefmark{3} and \v{C}edomir Stefanovi\'{c}\IEEEauthorrefmark{1}}
\IEEEauthorblockA{
\IEEEauthorrefmark{1}Department of Electronic Systems, Aalborg University, Denmark\\
\IEEEauthorrefmark{2}University of Novi Sad, Serbia \\
\IEEEauthorrefmark{3}German Aerospace Center, Munich, Germany\\
Email: hmf@es.aau.dk, dejanv@uns.ac.rs, andrea.munari@dlr.de, cs@es.aau.dk}
}

	\maketitle
	
	\begin{abstract}
Enabling real-time communication in Industrial Internet of Things (IIoT) networks is crucial to support autonomous, self-organized and re-configurable industrial automation for Industry 4.0 and the forthcoming Industry 5.0.  In this paper, we consider a SIC-assisted real-time IIoT network, in which sensor nodes generate reports according to an event-generation probability that is specific for the monitored phenomena.
 The reports are delivered over a block-fading channel to a common Access Point (AP) in slotted ALOHA fashion, which leverages the imbalances in the received powers among the contending users and applies successive interference cancellation (SIC) to decode user packets from the collisions.
 We provide an extensive analytical treatment of the setup, deriving the Age of Information (AoI), throughput and deadline violation probability, when the AP has access to both the perfect as well as the imperfect channel-state information.
 We show that adopting SIC improves all the performance parameters with respect to the standard slotted ALOHA, as well as to an age-dependent access method. The analytical results agree with the simulation based ones, demonstrating that investing in the SIC capability at the receiver enables this simple access method to support timely and efficient information delivery in IIoT networks.
	\end{abstract}
\section{Introduction}\label{sec:intro}
The Internet of Things (IoT) technology is evolving rapidly as a worldwide network of interconnected intelligent devices capable of sensing, communicating, and information processing to support a variety of applications~\cite{IoT}. As a subset of the IoT, Industrial IoT (IIoT) has gained significant attention as a key pillar of the era of Industry 4.0 and the forthcoming Industry 5.0 as well~\cite{industry-5.0}.
IIoT applications are more dependant on strict reliable and real-time requirements in order to maintain seamless and stable functionality of the industrial system~\cite{urllc}. 
A typical IIoT network is realized via Wireless Sensor Network (WSN) with a set of sensors deployed to monitor the environment and transmit the sensed information to a central Access Point (AP) for further processing. 
In event-driven IIoT networks, sensors are triggered by unpredictable events (e.g., alarms) to transmit status updates about the monitored physical phenomena. 
These updates are typically of high importance and should be delivered within a predefined deadline~\cite{deadline}, which can be understood as a packet-level metric that was in the initial focus of 5G URLLC standardization by 3GPP~\cite{3GPPURLLC}.
On the other hand, from the process-level perspective, the staleness of the collected updates is crucial for WSNs deployed in process control and monitoring scenarios as it affects the derived intelligent and autonomous decisions.

Recently, Age of Information (AoI) metric has emerged to characterize information freshness at the receiver side, defined as the time elapsed since the latest received update was generated~\cite{AoI}.
AoI can be understood as a process- (i.e., application-) level metric, which goes beyond the initial URLLC packet-level metrics. 
ALOHA is an appealing solution for medium-access control 
due to its simplicity and compatibility with the distributed nature of IIoT systems.
Specifically, ALOHA-based schemes flexibly cope with the uncertainty in the user activity patterns and the changes in the user population (i.e., users leaving and joining the network), in contrast to scheduled access schemes.
Their main downside is the performance sensitivity to collisions among transmitting users, which are unavoidable in this distributed and uncoordinated protocol. 
However, the performance of ALOHA-based schemes can be significantly improved by using advanced physical layer techniques that are able to retrieve the individual signals from collisions.  
In fact, such advanced access schemes in essence promote collisions, and are collectively referred to as non-orthogonal multiple access (NOMA) schemes \cite{NOMA_p0}.
Successive interference cancellation (SIC)~\cite{SIC} is a typical multi-user detection technique in NOMA where concurrent received signals are decoded in a descending order of their instantaneous received powers.

In this paper, we investigate a setup of a real-time IIoT in which a set of sensors perform event-driven reporting to a common AP using a simple slotted ALOHA scheme over a block-fading channel.
The AP is capable of advanced physical layer processing, leveraging the imbalances in the received powers among the contending users and applying SIC to decode user packets from the collisions.
 We perform an extensive analytical treatment of the setup and derive the AoI, throughput and deadline violation probability, which are process-level, access-network level and packet-level metrics, respectively, 
 both for the cases when the AP has access to the perfect and the imperfect channel-state information (CSI).
 We show that use of SIC improves all the performance parameters with respect to the standard slotted ALOHA, as well as to a recently reported age-dependent access method~\cite{dependant-AoI}.
 The derived analytical results agree with the simulation based ones, demonstrating that investing in the SIC capability at the receiver enables this simple scheme to support timely and throughput-efficient information delivery.

The rest of the paper is organized as follows.
Section~\ref{Related} presents the related work.
Section~\ref{system-model} elaborates the system model and
Section~\ref{analysis} presents the analysis.
Section~\ref{results} presents the performance evaluation.
Section~\ref{sec:conclusions} concludes the paper.

\section{Related Work}
\label{Related}

In \cite{schedule_1, schedule_2, schedule_4}, the authors propose different centralized scheduling protocols to minimize the AoI in IoT networks.
Although a promising solution to improve the AoI, the centralized scheduling approach is unsuitable to cope with non-trivial network dynamics when the number of active nodes change with time e.g., nodes joining or leaving the network or entering sleep mode to save energy.
Therefore, different decentralized, random access protocols are explored to improve the AoI performance in IoT networks \cite{ Random_2}.
In \cite{Random_2}, the authors propose a modification to the conventional slotted-ALOHA by introducing minislots to mitigate collisions and improve the network-wide AoI.
Age-dependant random access is proposed in \cite{dependant-AoI,dependant_1} where each node transmits with a certain probability only when its age exceeds a predefined threshold.
The mentioned works adopt the classical collision-channel model, neglecting the key effects of physical attributes in wireless systems, such as fading, path loss, and interference.
Moreover, these works mainly adopt the generate-at-will model for packet arrivals, which is not practically applicable in event-driven IIoT networks.
The authors in~\cite{NOMA_1} investigate the potential of NOMA in improving the AoI performance in a two-user network setup.
The work in~\cite{NOMA_2} extends the simple setup of~\cite{NOMA_1} to a multi-user system by dynamically switching between OMA and NOMA modes to minimize the AoI.
The focus of these works is mainly on the AoI performance without considering other metric such as throughput and delay, which are important in many event-driven IIoT applications.

\section{System Model}
\label{system-model}

We consider an event-driven IIoT network consisting of $N$ nodes with infinite buffers that transmit status updates to a single AP via a shared time-slotted channel.
On detection of an event, each node samples the underlying process, generates a status update, and contends to access the channel and to transmit the update with other active nodes in a slotted-ALOHA fashion.
For such setup, we model the packet arrivals as independent Bernoulli process in which an update arrives to the buffer of a node in each slot with probability $p_a$.
Each node attempts to transmit the update at the Head of the Line (HoL) with a probability $p$ in each slot following First In First Out (FIFO) discipline.
Successful transmissions are acknowledged via an error-free channel, and a failed update is retransmitted until either successfully delivered or its deadline expires; we assume that all updates have the same deadline $D$.
The considered setup is common to many IIoT applications, e.g., process monitoring and control in plastic extrusion industry~\cite{plastic}, where a set of sensor nodes are distributed to monitor the temperature profile along the barrel of the plastic extruder~\cite{mixed}.
Upon detecting a predefined deviation in the temperature profile, the nodes transmit updates to the AP, which analyses the updates and performs a temperature control mechanism. 

The AP uses the dynamic-ordered SIC~\cite{pathloss} to decode signals collided in a slot -- the decoding order is dynamically established and performed in the descending order of the instantaneous received powers of the collided transmissions. 
The composite signal received at the AP is given as
\begin{equation} \label{rece_signal}
y= \sum_{i\in\mathcal{T}}h_i\sqrt{P}x_i+w   
\end{equation}
where $\mathcal{T}$ is the set of active nodes in a given slot with $|\mathcal{T}| = m\leq N$; $x_i$ is the transmitted signal of user $n_i$, where $\mathbb E\{|x_i|^2=1\}$, $\forall n_i$; $P$ is the transmission power of $n_i$, which is equal for all nodes; and $w\sim \mathcal{CN}(0,\sigma^2)$ is the zero-mean additive white Gaussian noise with variance $\sigma^2$.
We denote $h_i= c_i /\sqrt{L_{d_i}} $ as the channel coefficient for the link between $n_i$ and the AP where $c_i\sim \mathcal{CN}(0,1)$ is the small-scale Raleigh fading coefficient following complex Gaussian distribution with zero mean and unit variance, and the large-scale fading component $L_{d_i}$ denotes the path loss for the node located at a distance $d_i$ from the AP, modelled by the free-space path loss model~\cite{pathloss}.
We consider a block Rayleigh-fading channel (i.e., the fading coefficient is constant during a slot) due to the fact that the transmitted signals from different nodes suffer from uncorrelated channel fading.
By $I_i = P|h_i|^2$ we denote the instantaneous power of the received signal, which is exponentially distributed with parameter $\lambda_i = 1/\E[I_i]$~\cite{pathloss}, i.e., its
probability distribution function (PDF) is given by
\begin{equation}\label{pdf_SIC}
f_{I_i}(r)=\lambda_i e^{-\lambda_i r}.   
\end{equation}

Considering that there are $1\leq m \leq N$ simultaneously transmitting nodes within the same slot, we define the outage probability $\Gamma_{i|m}$ as the probability that the received signal power of node $n_i$ is lower than the threshold required for the target transmission rate $\hat{R}$ (bits/s/Hz).
Assuming that only a single node, say $n_i$, is transmitting in a given slot (i.e., $m=1$), and defining $\gamma=2^{\hat{R}}-1$, the outage probability $\Gamma_{i|1}$ is 
\begin{equation}\label{outage}
  \Gamma_{i|1}= \textrm{Pr}\{I_i<\gamma \sigma^2\}=\int_0^{\gamma \sigma^2} \lambda_i e^{-\lambda_i I_i} \mathrm{d}I_i
  =1-e^{-\lambda_i\gamma \sigma^2}  .
\end{equation}

\section{Analysis of The Performance Metrics}
\label{analysis}

\subsection{Average AoI}

\begin{figure}[t!] 
		\centering
		\includegraphics[width= 1\linewidth]{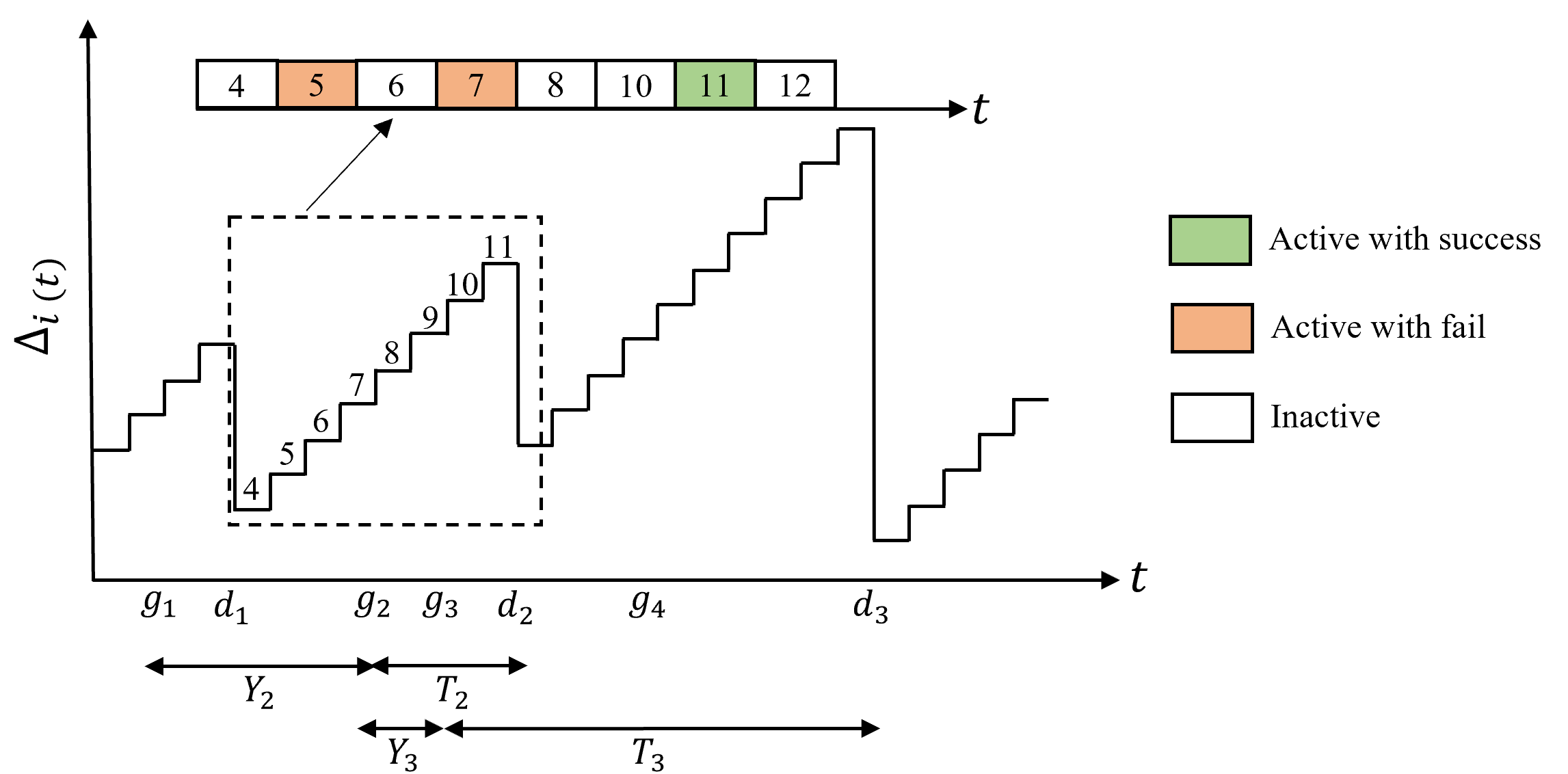}
		\caption{AoI evolution of arbitrary node $n_i$.  \label{AoI-ev}}
	\end{figure}

Without loss of generality, we analyze the average AoI for an arbitrary node $n_i$.
Fig.~\ref{AoI-ev} shows an example of the evolution of the AoI of $n_i$, where $g_k$ is the slot index of the generated $k$th update and $d_k$ is the slot index at which it was successfully received by the AP.
The instantaneous AoI $\Delta_i(t)$ is recorded at the end of each slot $t=1, 2, 3, \dots$
Assuming that the $k$th update is the one recently received, then the evolution of the instantaneous AoI can be expressed by 
\begin{equation}
\Delta_i (t+1)=\begin{cases}
\Delta_i(t)+1 & \text{failed transmission}\\
(t-g_k)+1 & \text{successful transmission}.
\end{cases}
\end{equation}
Denoting the system time and the interarrival time of the $k$th update as $T_k = d_k-g_k$ and $Y_k = g_k-g_{k-1}$, respectively, the average AoI $\overline{\Delta}_i$ of $n_i$ is given by
\begin{equation}
\begin{split}
\label{eq:aAoI}
 \overline{\Delta}_i &= p_a \left(\E[YT]+\frac{\E[Y^2]}{2}+\frac{\E[Y]}{2}\right) \\    
 & = \frac{1}{p_a} + \frac{1-p_a}{q_s-p_a}+\frac{p_a}{q_s}-\frac{p_a}{q_s^2} 
\end{split}
\end{equation}
where $q_s$ is the successful update probability, i.e., the probability that $n_i$ successfully delivers the HoL update to the AP, derived in Section~\ref{sec:SUP}. The detailed derivations of \eqref{eq:aAoI} are omitted due to the limited space, and follow the same procedures as in \cite{AoI_model}.
Henceforth, we drop the subscript $i$ in $\overline{\Delta}_i$, as its value is the same for all nodes, as shown by \eqref{eq:aAoI}.

\setcounter{equation}{11}
	\newcounter{storeeqcounter}
	\newcounter{tempeqcounter}
	\addtocounter{equation}{1}%
	\setcounter{storeeqcounter}%
	{\value{equation}}
	\begin{figure*}[!t]
		\normalsize 
		\setcounter{tempeqcounter}{\value{equation}} 
		
		\begin{IEEEeqnarray}{rCl}  
			\setcounter{equation}{\value{storeeqcounter}} 
			\begin{split} \label{avg-outage}
			\overline{\Gamma}_{i|m}& = \frac{1}{\binom{N-1}{m-1}} \sum_{{\underset{i = \pi_i}{\pi\in \mathcal{O}_m}}}\Gamma_{i|m} \\  
			& = \frac{1}{\binom{N-1}{m-1}} \sum_{{\underset{i = \pi_i}{\pi\in  \mathcal{O}_m}}} \sum_{{\underset{i = \pi_i}{\pi(m)\in \mathcal{O}_m}}}\frac{\prod_{i=2}^m\lambda_{\pi_i}}{\prod_{i=2}^m\left(\sum_{k=1}^i \lambda_{\pi_k}\right)}\left(1-\prod_{l=1}^m \mathrm{Pr}\left\{\textrm{log} \left(1+\frac{I_i}{\sigma^2+\sum_{j=i+1}^{m}I_j}\right)\geq \hat{R}|\pi(m)\right\}\right)
			\end{split}
		\end{IEEEeqnarray}
		\setcounter{equation}{\value{tempeqcounter}} 
		\hrulefill
	\end{figure*}
	\setcounter{equation}{5}

\subsection{Successful Update Probability}
\label{sec:SUP}

With the dynamic-ordered SIC scheme \cite{pathloss}, the AP tries to decode collided signals individually following a decoding order $\pi(m) = \{\pi_1, \pi_2, ..... \pi_{m}\}$ which represents an ordered set of the indices of $m$ contending nodes in a given timeslot with $I_{\pi_1}> I_{\pi_2}, ..... >I_{\pi_m}$.
The AP first tries to decode the signal with the strongest instantaneous power $I_1$ while treating other signals as interference.
If $\pi_1$ is successfully decoded, the AP then removes the reconstructed signal from the composite received signal and tries to decode $\pi_2$, etc.
The process continues until all the $m$ nodes are decoded successfully or an outage event is declared.
Upon finding the signal $\pi_i$ in outage, the AP subsequently fails to decode the remaining $m-i$ signals, and the backlogged nodes attempt to retransmit in the next timeslot with probability $p$.   
 
Based on \eqref{pdf_SIC}, the probability of the decoding order $\pi(m)$ with $I_{\pi_1}> I_{\pi_2}, ..... >I_{\pi_m}$ is given as
\begin{equation}\label{order_pr}
\begin{split}
 &\mathrm{Pr}\{\pi(m)\}  = \mathrm{Pr}\{I_1> I_2, ..... >I_m\}\\
 & = \int_0^{\infty} \lambda_{\pi_m} e^{-\lambda_{\pi_m} I_{\pi_m}} \textrm{d} I_{\pi_m} .....\, \int_{I_{\pi_3}}^{\infty} \lambda_{\pi_2} e^{-\lambda_{\pi_2} I_{\pi_2}} \textrm{d} I_{\pi_2} \\
 & \times \int_{I_{\pi_2}}^{\infty} \lambda_{\pi_1} e^{-\lambda_{\pi_1} I_{\pi_1}} \textrm{d} I_{\pi_1}  = \frac{\prod_{i=2}^m\lambda_{\pi_m}}{\prod_{i=2}^m\left(\sum_{j=1}^i \lambda_{\pi_j}\right)}.
\end{split}
\end{equation}
Under the decoding order $\pi (m)$, the achievable rate $R_i$ of node $n_i$ when contending with $m$ nodes in the given slot is
\begin{equation} \label{rate_ni}
R_i = \textrm{log} \left(1+\frac{I_i}{\sigma^2+\sum_{j=i+1}^{m}I_j}\right).
\end{equation}
To successfully decode the status update from $n_i$, it must be $R_i \geq \hat{R}$. In addition, the AP must have decoded all the prior $(i-1)$ nodes.
Hence, the event $S_{i|\pi}^m$, representing the successful delivery of the update of $n_i$ under $\pi(m)$, is expressed by
\begin{equation}\label{sucess_event}
\mathrm{Pr}\{S_{i|\pi}^m\}= \mathrm{Pr}\{R_i \geq \hat{R}|\pi(m)\}.
\end{equation}
The outage probability of $n_i$ given $\pi(m)$ is
\begin{equation}
 \Gamma_{i|m}^{\pi(m)} = 1- \mathrm{Pr}\{S_{1|\pi}^m \cap S_{2|\pi}^m \cap ... \cap S_{i|\pi}^m\} = 1-\prod_{l=1}^i \mathrm{Pr}\{S_{l|\pi}^m\}
\end{equation}
where the events $S_{x|\pi}^m$ and $S_{y|\pi}^m$ ($x \neq y$) are mutually independent. In the dynamic-ordered SIC, the decoding order dynamically varies due to the channel conditions, hence the outage probability of $n_i$ can be obtained by averaging over all possible decoding orders $\mathcal{O}_m$  as follows 
\begin{equation}\label{outage_final}
 \Gamma_{i|m}  = \E_{\mathcal{O}_m} \left[ \Gamma_{i|m}^{\pi(m)} \right] =  \sum_{\pi(m)\in \mathcal{O}_m, i = \pi_i} \mathrm{Pr}\{\pi(m)\} \Gamma_{i|m}^{\pi(m)}.
\end{equation}
Based on \eqref{outage_final}, the successful update probability $q_s$ becomes
\begin{equation}
\begin{split}
q_s = \sum_{m=1}^N \binom{N-1}{m-1} p^{m} (1-p)^{n-m} (1-\overline{\Gamma}_{i|m})
\end{split}
\end{equation}
where $\overline{\Gamma}_{i|m}$ is given in~\eqref{avg-outage}. 	
To further simplify this expression, we consider that $\lambda_{\pi_i} = \lambda,\, \forall i$ (i.e., all nodes have equal average SNR).
Then, from \eqref{order_pr}, we have $\mathrm{Pr}\{\pi(m)\} = \frac{1}{m!}$ and \setcounter{equation}{12}
\begin{equation}\label{suc_decod}
\mathrm{Pr}\{S_{i|\pi}^m\} = \binom{m}{i}\frac{e^{-i\lambda \gamma\sigma^2}}{(\gamma i + 1)^{(m-i)}}.    
\end{equation}
The successful decoding probability $\Gamma_{i|m}$ in \eqref{outage_final} can be now denoted as $\Gamma_{m}$ and is given by
\begin{equation}\label{outage_finall}
\begin{split}
 \Gamma_{m} &=  \sum_{\pi(m)\in \mathcal{O}_m, i = \pi_i} \mathrm{Pr}\{\pi(m)\} \Gamma_{i|m}^{\pi(m)}  \\
 & =  \mathrm{Pr}\{\pi(m)\} (m-1)! \sum_{i=1}^m \Gamma_{i|m}^{\pi(m)}\\
 & = 1-\frac{1}{m} \sum_{i=1}^m \left[\prod_{j=1}^i\binom{m}{j}\frac{e^{-j\lambda \gamma\sigma^2}}{(\gamma j + 1)^{(m-j)}}\right].
\end{split}
\end{equation}
Accordingly, $q_s$ becomes
\begin{equation}\label{success1}
q_s = \sum_{m=1}^N \binom{N-1}{m-1} p^{m} (1-p)^{N-m} (1-\Gamma_{m}).
\end{equation}
\subsection{Throughput}
We define the throughput $S_\text{th}$ as the mean number of successfully decoded packets per slot.
With the dynamic-ordered SIC, having the $i$th signal decoded successfully means that the previous ($i-1$) signals have been already decoded.
Hence, from \eqref{suc_decod}, $\mathrm{Pr}\{S_{i|\pi}^m\}$ implies that there are at least $i$ successfully decoded signals in a given slot.
Then, we express $ P^{m}_{i,\text{SIC}}$, which is the probability of decoding at least $1\leq i \leq N$ signals while there are $m$ out of $N$ signals transmitting, by
\begin{equation}\label{SIC_m}
  P^{m}_{i,\text{SIC}} = \prod_{j=1}^i \left[ \sum_{m=j}^N \binom{N}{m} p^m (1-p)^{(N-m)} X(j)\right]
\end{equation}
where $X(j) = \binom{m}{j}\frac{e^{-j\lambda \gamma\sigma^2}}{(\gamma j + 1)^{(m-j)}}$.
The probability of decoding exactly $i$ signals, $1\leq i \leq N-1$, can be calculated as
\begin{equation}
  P_{i,\text{SIC}} =  
\begin{cases}
P^{m}_{i,\text{SIC}} - P^{m}_{i+1,\text{SIC}} & 1\leq i \leq N\\
P^{m}_{N ,\text{SIC}} & i = N
\end{cases}
\end{equation}
Finally, the throughput $S_\text{th}$ is given as
\begin{equation}
S_\text{th} = \E[i] = \sum_{i=1}^N i \,  P_{i,\text{SIC}} \,. 
\end{equation}
\subsection{Deadline Violation Probability}
\begin{figure}[t!] 
		\centering
		\includegraphics[width= 0.8\linewidth]{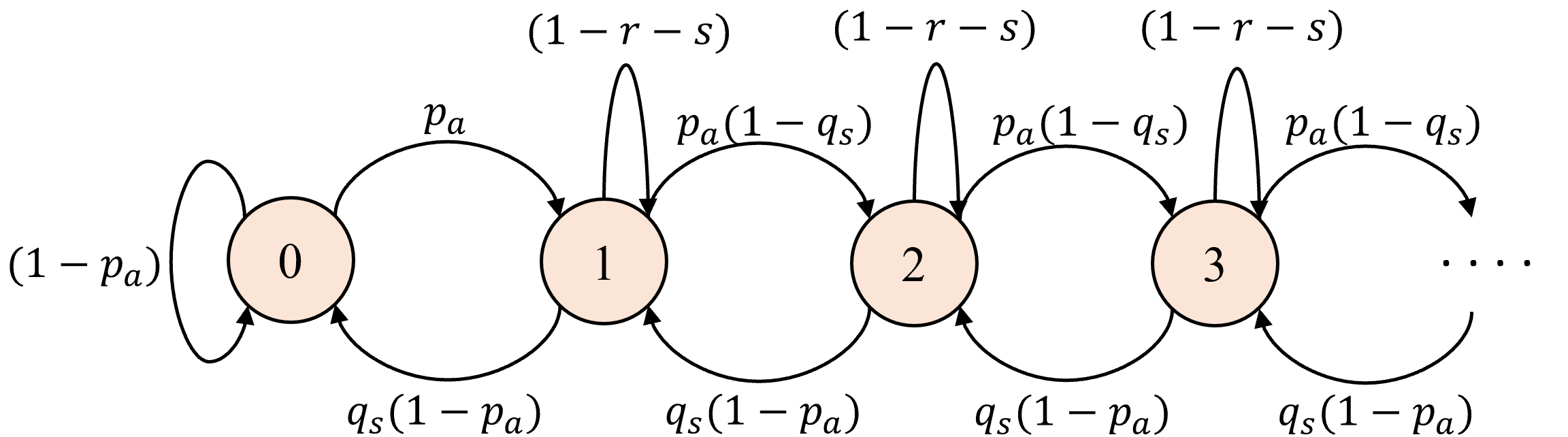}
		\caption{DTMC model of the queue size of node $n_i$.  \label{DTMC-deadline}}
	\end{figure}

In IIoT networks, the AP utilizes the collected updates for further processing and control actions.
A received update is regarded as valid if its delivered within a predefined deadline since its generation, otherwise it becomes useless to the system.
Here we derive the deadline violation probability $P_D$, which is the probability that the transmission delay exceeds a predefined deadline $D$, i.e., $P_D =\mathrm{Pr}\{T>D\}$.
In order to calculate $P_D$, we derive the distribution of the steady-state delay $T$ of an update.
Each node attempts to transmit the HoL update until its successfully delivered with probability $q_s$.
Hence, the service time $A_j$ of the $j$th update is a geometrically distributed random variable (RV) with parameter $q_s$.
Consider a tagged update that arrives while there are $K$ updates already in the buffer.
The delay of this update is the random sum $T = A_1 + A_2 + ... + A_K$, where $A_j,\, j = 1, 2, ...., K,$ are independent and identically distributed geometric RVs with parameter $q_s$.
In order to find the PDF of $T$, we first calculate its probability generating function $G_T(z)$.
The Discrete-Time Markov Chain (DTMC) in Fig.~\ref{DTMC-deadline} shows the queue evolution of $n_i$, where $r= p_a(1-q_s)$ and $s = q_s(1-p_a)$.
Let $Q_j$ be the steady-state probability of having $j$ updates in the queue of $n_i$, then using the balance equations of the DTMC in Fig. \ref{DTMC-deadline}, we obtain the following 
\begin{equation}
Q_j = 
    \begin{cases}
\rho^{j-1} Q_1 & j\geq 1\\
\frac{q_s (1-p_a)}{p_a} Q_1 & j=0
    \end{cases}
\end{equation}
where $\rho = \frac{r}{s}$, $Q_1 = \frac{p_a(1-\rho)}{q_s}$. according to the convolution property of their
generating functions~\cite{conv}, the probability generating function $G_T(z)$ is
\begin{equation}\label{PGF}
\begin{split}
 G_T(z) &= \sum_{j=1}^\infty \left(\frac{q_s z}{1-(1-q_s)z}\right)^j (1-\rho)\rho^{(j-1)}  \\
 & = \frac{q_s (1-\rho)z}{1-(1-q_s(1-\rho))z}.
\end{split}
\end{equation}
Based on \eqref{PGF}, the PDF of $T$ is obtained as
\begin{equation}\label{PDF_T}
f_T(t) = q_s (1-\rho)(1-(q_s(1-\rho)))^{(t-1)}
\end{equation}
and the deadline violation probability $P_D$ is obtained as
\begin{equation}
\begin{split}
P_D &= \mathrm{Pr}\{T>D\} = 1-\mathrm{Pr}\{T\leq D\} \\
 &= 1-\sum_{i=1}^D f_T(i) = (1-(q_s(1-\rho)))^D.
\end{split}
\end{equation}


\subsection{The effect of Imperfect CSI}\label{CSI_analysis}
So far, we considered that the AP maintains perfect CSI to successfully perform SIC on the collided signals.
Practically, channel estimation errors are inevitable as channel estimation with high accuracy imposes significant system overhead~\cite{CSI_im}.
With imperfect CSI, the AP may adopt incorrect decoding order when the estimated instantaneous received power fails to reflect the actual one.
In addition, such estimation error introduces extra interference to the SIC process, which in turn increases the outage probability.
In the following, we focus on the effect of extra interference where the effect of the change in the decoding order is small and can be neglected. Let $\hat{h}_i$ be the estimated channel coefficient with estimation error $\epsilon_i$, then the fading coefficient $h_i$ in \eqref{rece_signal} is given as
\begin{equation}
    h_i = \hat{h}_i+\epsilon_i,
\end{equation}
where $\epsilon_i\sim \mathcal{CN}(0,\sigma_{\epsilon}^2)$.
Hence, the estimated received power $\hat{I}_i = P|\hat{h}_i|^2$ is exponentially-distributed with parameter $\hat{\lambda}_i$.
In this case, the interference to node $n_i$ is $\sum_{j=i+1}^{m}I_j + \sum_{l=1}^i \phi_l$, where $\phi_l$ is the difference between the received signal power and its estimation, which is exponentially distributed with parameter $v_i$.
With the assumption of equal SNR for all nodes, the outage probability $\hat{\Gamma}_{m}$ becomes
\begin{equation}\label{outage-CSI}
   \hat{\Gamma}_{m}= 1-\frac{1}{m} \sum_{i=1}^m \left[\prod_{j=1}^i\binom{m}{j}\frac{\textrm{exp}\left({-j\lambda \gamma(\sigma^2+\frac{j}{v})}\right)}{(\gamma j + 1)^{(m-j)}}\right].
\end{equation}
Compared to \eqref{outage_finall}, the additional interference component $\frac{j}{v}$ due to the imperfect CSI clearly affects the outage probability.
Since the  successful update probability $q_s$ depends on $\hat{\Gamma}_{m}$, the imperfect CSI will in turn have effect on AoI, $S_\text{th}$ and $P_D$.

    \begin{figure}[t!] 
		\centering
		\includegraphics[width= 0.8\linewidth]{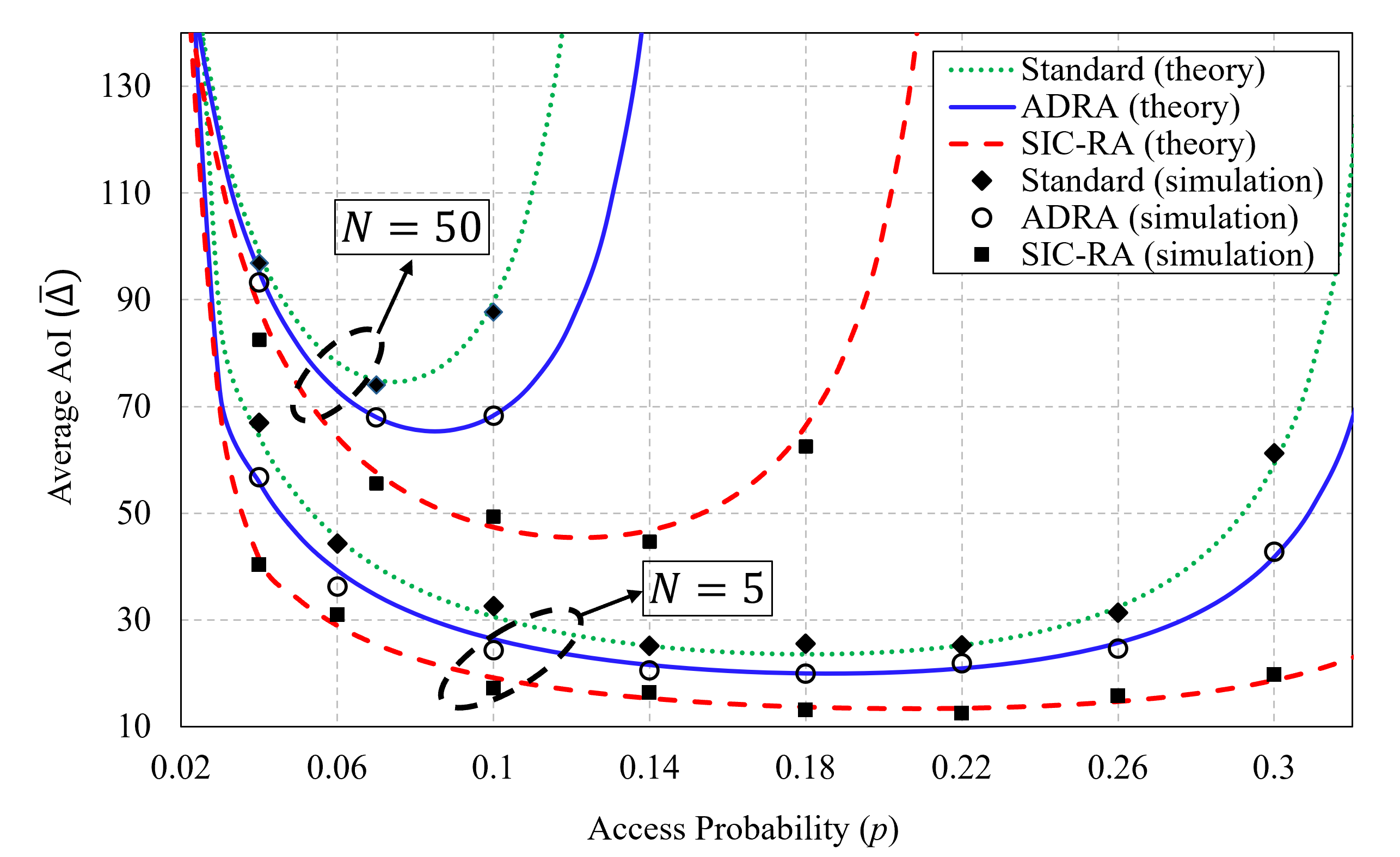}
		\caption{Evaluation of the average AoI against the transmission probability $p$ with $N\in \{5,50\}$.  \label{AoI-p}}
	\end{figure}
    \begin{figure}[t!] 
		\centering
		\includegraphics[width= 0.8\linewidth]{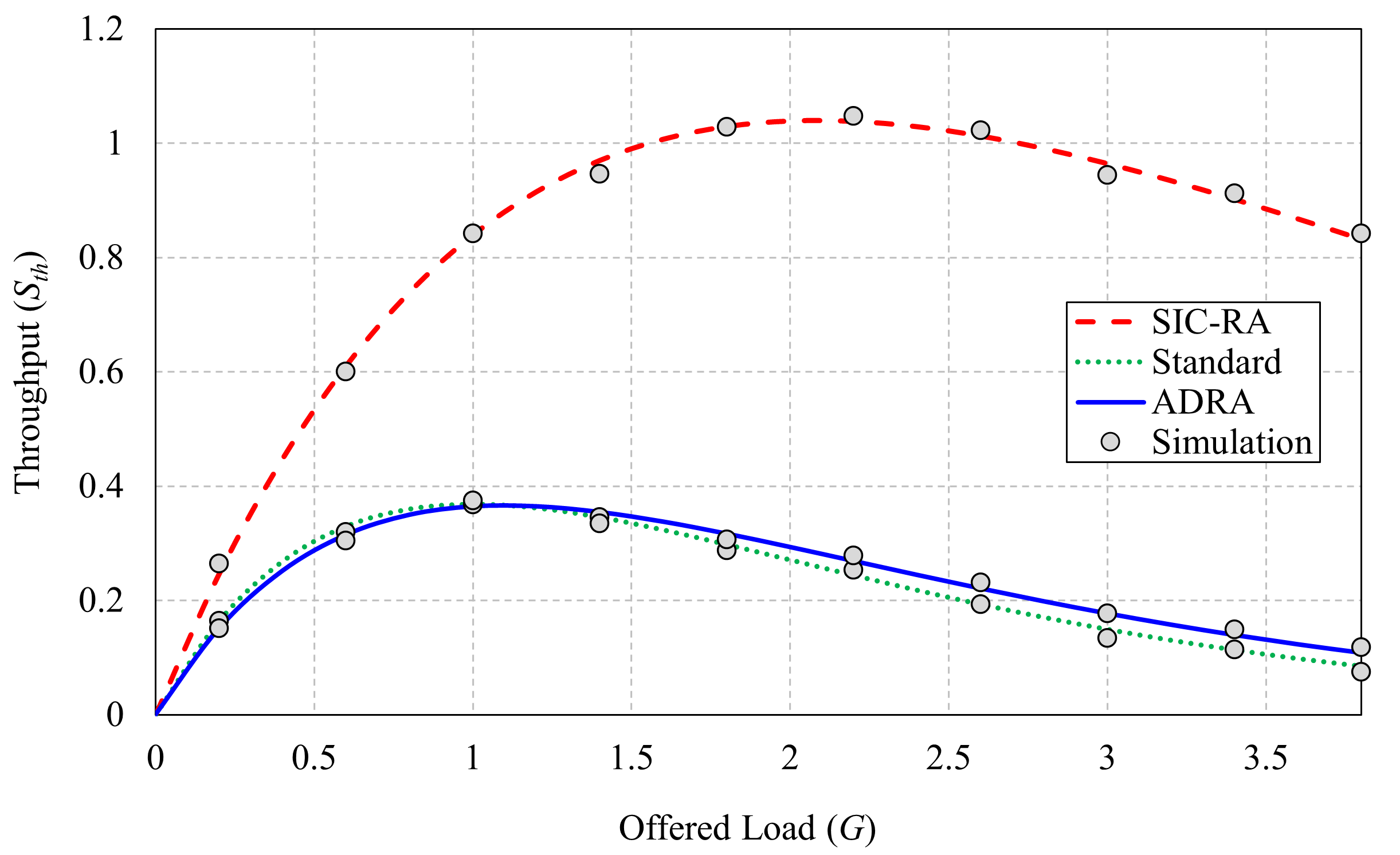}
		\caption{Throughput comparison with varying offered load $G$.  \label{th-G}}
	\end{figure}
    \begin{figure}[t!] 
		\centering
		\includegraphics[width= 0.8\linewidth]{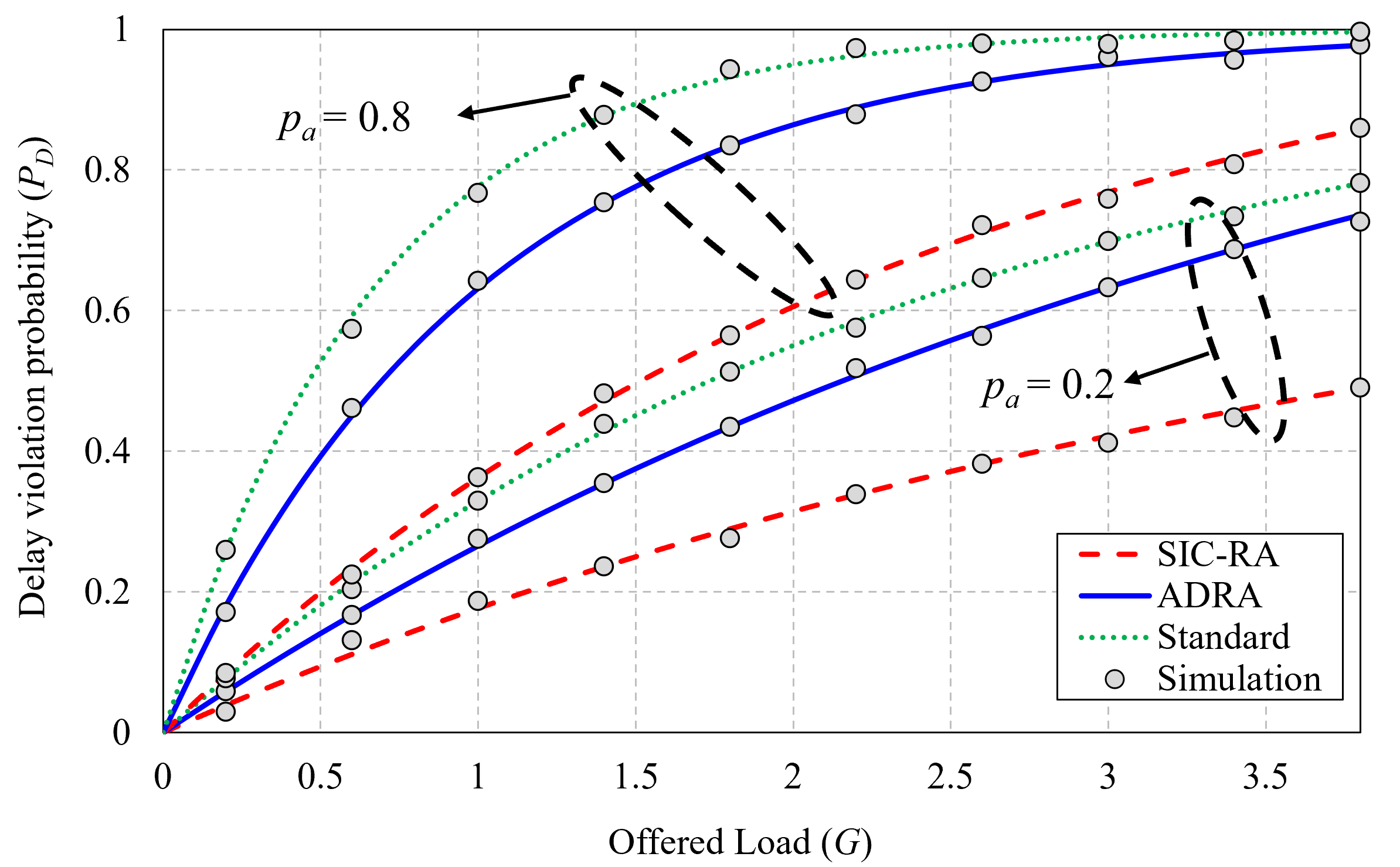}
		\caption{Comparison of the deadline violation probability under varying offered load $G$ with $p_a\in \{0.2,0.8\}$.  \label{P_D}}
  
	\end{figure}


\section{Results and Discussion}
\label{results}
In this section, we evaluate the system performance in terms of the average AoI, throughput and deadline violation probability. 
In addition, we validate our analysis in Section~\ref{analysis} via discrete-event simulations in MATLAB.
The simulation setup consists of a network of $N$ nodes randomly distributed in a $200\,\textrm{m}\times 200\,\textrm{m}$ area.
We set the transmission power $P$ for all nodes to $20$~dBm and the arrival probability $p_a$ to 0.4.
The following results are averaged over 10 simulation runs, with each run lasting $10^5$ slots.
We also compare our results, referred to as SIC-RA, with the standard slotted ALOHA (referred to as Standard) and the other is the age-dependant random access in~\cite{dependant-AoI} (referred to as ADRA).

Fig.~\ref{AoI-p} compares the average AoI $\overline{\Delta}$ for SIC-RA, Standard and ADRA with varying $p$.
Clearly, the analytical results well match the ones obtained via simulations, validating our analysis in Section~\ref{analysis}.
As $p$ increases, $\overline{\Delta}$ first decreases as the nodes tend to transmit more often, and then, after some critical value of $p$ is surpassed, $\overline{\Delta}$ increases due to excessive collisions, prolonged queuing and consequential packet drops. 
SIC-RA achieves the lowest $\overline{\Delta}$ due to the adopted SIC functionality at the AP, which significantly increases the success probability compared to Standard and ADRA.
Further, SIC-RA attains a wider region with favorable $\overline{\Delta}$. 
The improvement gap increases when the network size grows from $N = 5$ to $N = 50$, demonstrating the benefits of collision resolution by the adopted SIC functionality.
For instance, at $p= 0.14$, or $N = 5$ SIC-RA improves $\overline{\Delta}$ compared to Standard and ADRA with $25\%$ and $38\%$, respectively, while for $N = 50$ the improvement increases to $71\%$ and $89\%$, respectively.
It is interesting to observe that the average AoI of SIC-RA for $N=50$ and a range of values of $p$ is lower than 50 slots, which is the minimum value for a round-robin based scheme.

Fig.\ref{th-G} compares the throughput $S_\text{th}$ of SIC-RA, Standard and ADRA with varying offered load, defined as $G=Np$. Evidently, SIC-RA achieves higher throughput than ADRA and Standard as the SIC feature enables efficient utilization of the channel via decoding and reception of multiple packets within a single slot.
We observe that ADRA maintains almost the same throughput performance as Standard which is upper bounded by $e^{-1}$.
In other words the age-dependant random access in ADRA does not have effect on improving the number of successful transmissions per slot.
From Fig.~\ref{th-G}, we can see that SIC-RA increases the maximum throughput in comparison to Standard and ADRA by almost $188\%$. 

Fig.~\ref{P_D} how the deadline violation probability $P_D$ depends on the offered load $G$, with a deadline of $D = 5$ slots and $p_a \in \{0.2,0.8\}$.
Clearly, SIC-RA achieves the lowest $P_D$, 
and the improvement gap gets increased as $p_a$ increases.
In ADRA, a node that has just delivered an update will refrain from transmitting until its instantaneous AoI exceeds some threshold.
Hence, backlogged updates at the output buffer will suffer extended queuing times, which is the dominant component in the steady-state delay $T$ and which degrades $P_D$.
In SIC-RA, the steady-state delay is significantly decreased and the backlogged updates have a higher chance to be delivered within the deadline.
Therefore, while ADRA favors the AoI over the delay performance, SIC-RA has the potential to improve both.
For instance, SIC-RA reduces $P_D$ by $44\%$ and $35\%$ compared to Standard and ADRA, respectively, with $G=1$ and $p_a = 0.2$, while the reduction is increased to $54\%$ and $48\%$, respectively, for $p_a = 0.8$ and the same $G$.

\begin{figure}[t!]
    \centering
    \begin{subfigure}[t]{0.5\textwidth}
        \includegraphics[width= 0.8\linewidth]{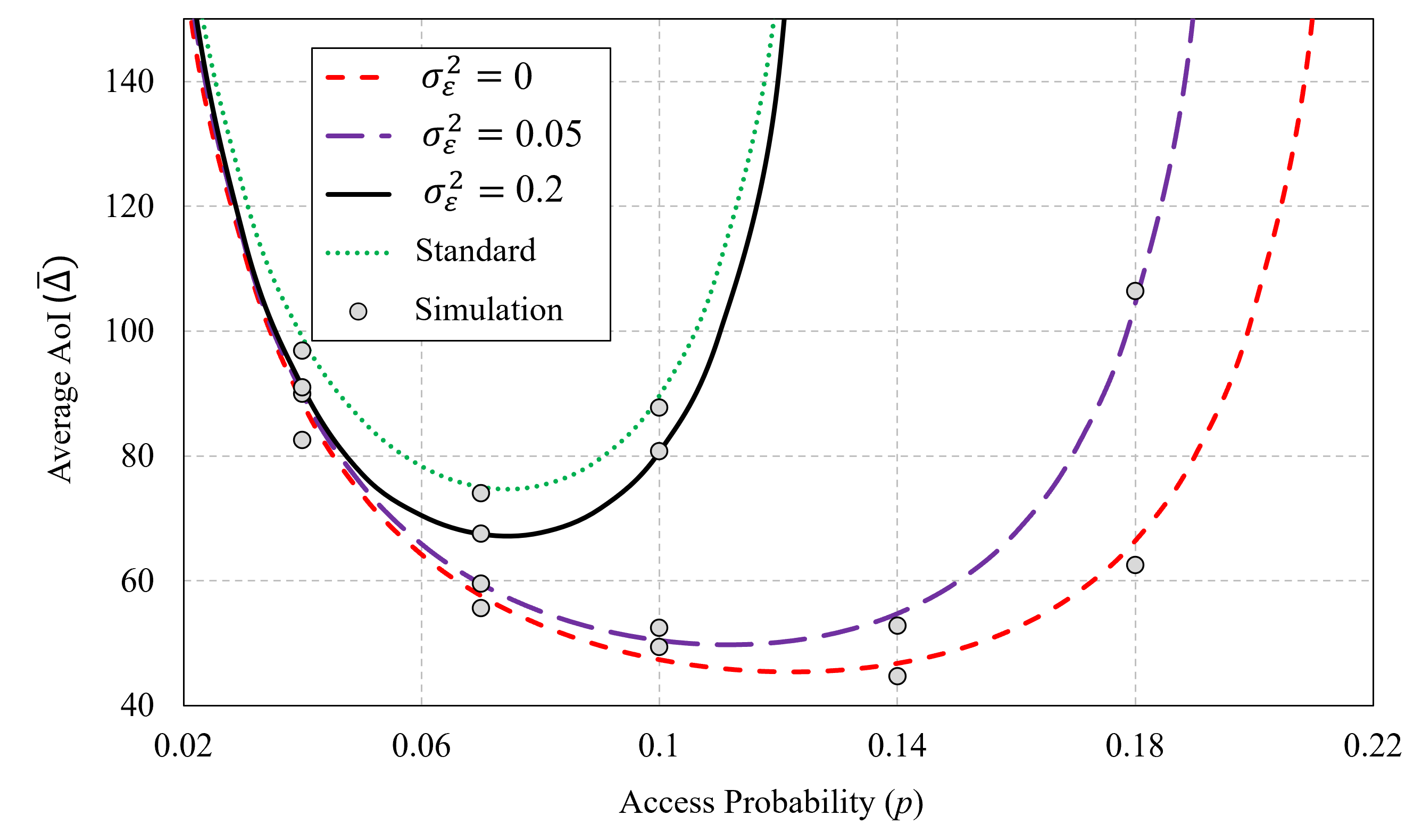}
        \caption{Average AoI with $N = 40$.}
    \end{subfigure}
    \begin{subfigure}[t]{0.5\textwidth}
        \includegraphics[width= 0.8\linewidth]{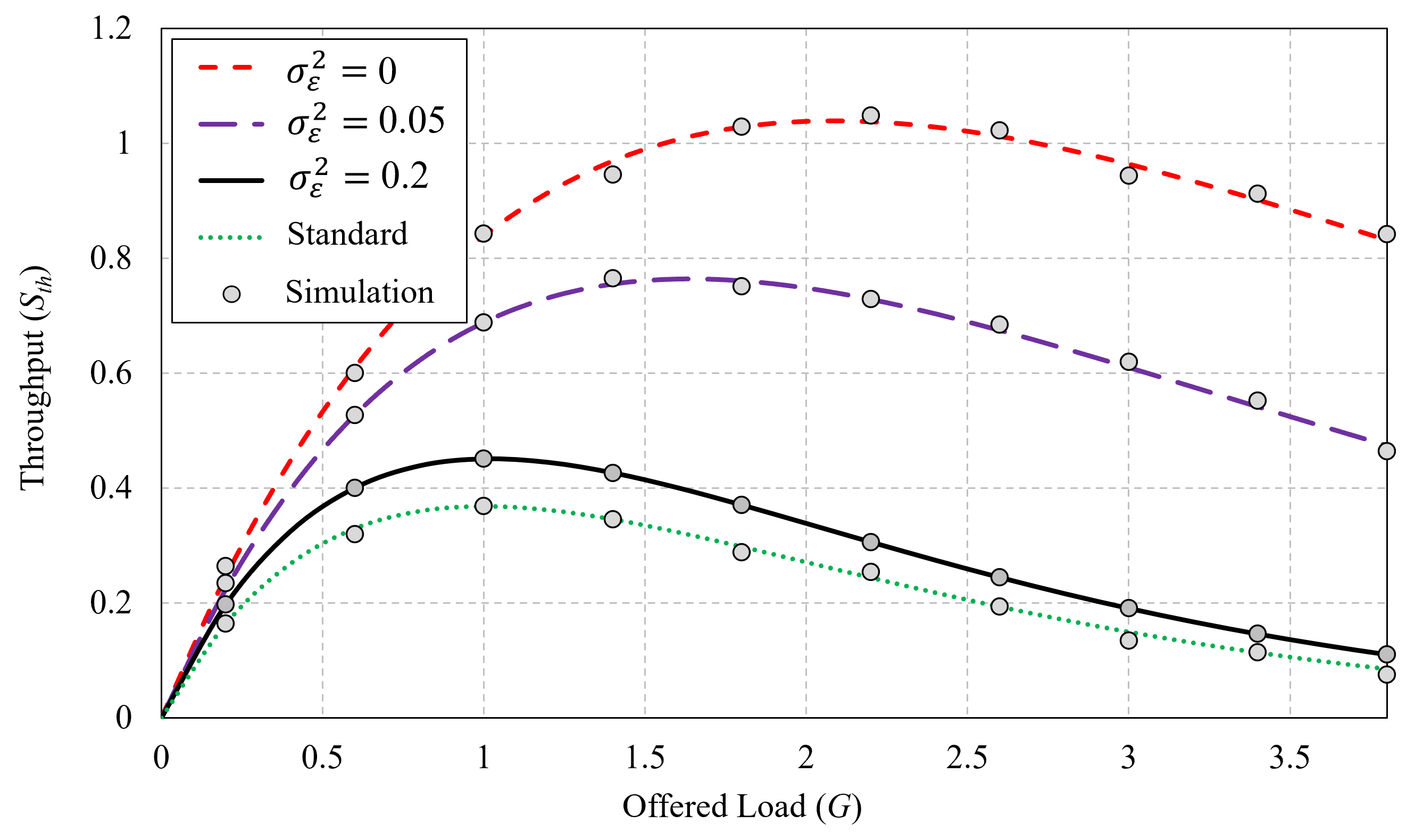}
        \caption{Throughput.}
    \end{subfigure}
    \begin{subfigure}[t]{0.5\textwidth}
        \includegraphics[width= 0.8\linewidth]{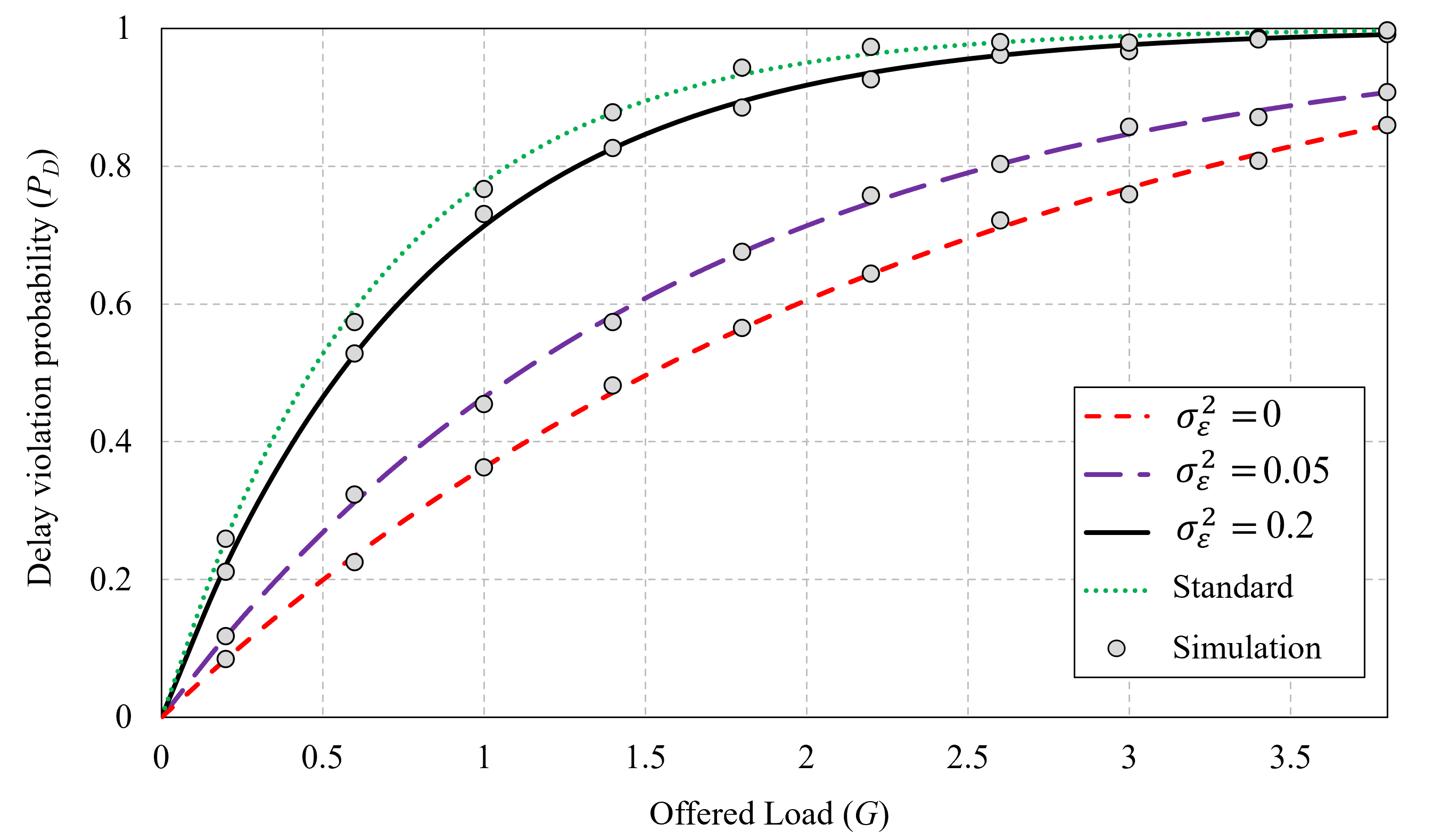}
        \caption{Deadline violation probability with $p_a = 0.8$.}
    \end{subfigure}
    \caption{Effect of imperfect CSI}\label{CSI}

\end{figure}

Finally, Fig.~\ref{CSI} shows the effect of imperfect CSI on the performance parameters, assuming the variance of the estimation error $\sigma^2_{\epsilon} = \{0,0.05, 0.2\}$ ($\sigma^2_{\epsilon} = 0$ corresponds to perfect CSI).
As shown by Fig.~\ref{CSI}, $\overline{\Delta}$, $S_{th}$ and $P_D$ become progressively degraded as $\sigma^2_{\epsilon}$ increases, which is due to the extra interference that increases the outage probability, as indicated by~\eqref{outage-CSI}.
With $\sigma^2_{\epsilon} = 0.05$, the decrease in the average AoI is almost $18\%$ at $p=0.1$, while the decreases in $S_\text{th}$ and $P_D$ are $25\%$ and $22\%$, respectively, at $G = 1.5$.
The performance is further compromised with the increase of $\sigma^2_{\epsilon}$, tending to the performance of standard slotted ALOHA. 

\section{Conclusion}
\label{sec:conclusions}

In this paper, we analyzed an event-driven reporting real-time IIoT scenario. In the considered system model,  the reporting nodes use a slotted ALOHA scheme, while the AP exploits channel-induced power imbalances to perform SIC based decoding of transmitted updates. The scheme shows favorable performance in terms of the average AoI, packet deadline violation probability and throughput, while not requiring implementation of an involved channel access policy by the users.
We also characterized the impact of the imperfect CSI on the performance parameters, showing that the AP should have a fairly precise estimates of the CSI in order to fulfill the scheme's potential.

\section*{Acknowledgement}
This paper has received funding from the European Union’s Horizon 2020 research and innovation programme under Grant Agreement number 856967. Andrea Munari acknowledges the support of the Federal Ministry of Education and Research of Germany in the programme of “Souverän. Digital. Vernetzt.” Joint project 6G-RIC, project identification number: 16KISK022.

	\bibliographystyle{IEEEtran}
\bibliography{main}
	
\end{document}